\documentclass[amssymb,aps,pra,reprint,groupedaddress,superscriptaddress,floatfix,showpacs]{revtex4-2}
\usepackage{graphicx}
\usepackage{dcolumn}
\usepackage{bm}
\usepackage[linkcolor=blue]{hyperref}
\usepackage[T1]{fontenc}
	
\begin{document}

\title{Pressure-induced toroidal order in molecular Sn$_2$P$_2$S$_6$ ferroelectrics}

\author{Konstantin Z. Rushchanskii}
\email[Corresponding author: ]{rushchanskii@gmail.com}
 \affiliation{Peter Gr\"{u}nberg Institut and Institute for Advanced Simulation, Forschungszentrum J\"{u}lich and JARA, 52425 J\"{u}lich, Germany}

\author{Mykola Medulych}
 \affiliation{Institute for Solid State Physics and Chemistry, Uzhhorod National University, 88000 Uzhhorod, Ukraine}
 
 \author{Vitalii Liubachko}
 \email[Corresponding author: ]{vitalii.liubachko@uzhnu.edu.ua}
 \affiliation{Institute for Solid State Physics and Chemistry, Uzhhorod National University, 88000 Uzhhorod, Ukraine}

\author{Yulian M. Vysochanskii}
 \email[Corresponding author: ]{vysochanskii@gmail.com}
\affiliation{Institute for Solid State Physics and Chemistry, Uzhhorod National University, 88000 Uzhhorod, Ukraine}%

\date{\today}

\begin{abstract}
For the Sn$_2$P$_2$S$_6$ crystalline compound with a rich temperature--pressure phase diagram that contains the ferroelectric phase, semiconductor-to-metal transition, and superconductive state, \mbox{\textit{ab initio}} molecular dynamics calculations were performed on a single unit cell and on a superstructure using an evolutionary algorithm in combination with density functional theory calculations to study structural transformations resulting from rearrangements in chemical bonds under pressure. The structure models of the pressure-induced lowest energy phases in the molecular Sn$_2$P$_2$S$_6$ crystal demonstrate the possibility of space-arranged chains or rings of P$_2$S$_6$ molecules. Pressure can break P-P bonds of P$_2$S$_6$ molecules, causing one of the phosphorous atoms to displace into the sulfur-formed octahedron and providing another phosphorous atom into an octahedral coordination with two tin atoms and four sulfur atoms. The results of the modeling correlate with the gradual rise of deviation of the pressure dependence of the unit cell volume of the Sn$_2$P$_2$S$_6$ structure, as determined by X-ray diffraction experiments, from the dependence calculated with the Birch-Murnaghan equation of state. It also reveals toroidally ordered local dipoles in the centrosymmetric phase, which may be related to the previously observed superconductivity in the high-pressure metallic phase of Sn$_2$P$_2$S$_6$.
\end{abstract}

\pacs{77.84.-s, 77.65.-j, 81.05.Zx, 61.50.Ah, 61.72.Bb, 61.72.Ji, 64.30.+t}

\maketitle

\section{Introduction}

The phosphorus chalcogenides of MM$'$P$_2$X$_6$ family (where M and M$'$ denote transition metals or elements of main group, X stands for S, Se) consists of the crystal compounds with bulk, layered, and chain crystal lattices, that are built by the frame of [P$_2$S(Se)$_6$]$^{4-}$ molecular anions joined into periodic lattices with different (i.e., 3D, 2D and 1D) dimensionality \cite{brec1986review, samal2021, susner2017, carpentier1974}. These compounds exhibit ferroicity and multiferroicity with different types of electric dipoles ordering (ferroelectric, ferrielectric, antiferroelectric or incommensurately modulated \cite{liubachko2020, maisonneuve1997, dziaugys2020}), and with several types of antiferromagnetic ordering \cite{ouvrard1985}. Also, the relaxor or dipole glass states can be found on the temperature--composition phase diagrams, as in the case of CuInP$_2$(Se$_x$S$_{1-x}$)$_6$ mixed crystals \cite{macutkevic2008}. In the Cu(Cr$_x$In$_{1-x}$)P$_2$S$_6$ solid solutions one can reach the multiglass state with simultaneous ``freezing'' of partially disordered antiferromagnetic and ferrielectric types \cite{kleemann2011}. In the case of Mn$_2$P$_2$S$_6$,  magnon-phonon interaction reflects the changes in the properties of the density of states caused by the phonon excitations in the low-dimensional crystal \cite{peschanskii2019}.

The ferroicity of M$_2$P$_2$X$_6$ compounds interacts with their transition from an insulating or semiconducting state to a metallic state. The layered van der Waals (vdW) Fe$_2$P$_2$S$_6$, Mn$_2$P$_2$S$_6$ and Ni$_2$P$_2$S$_6$ crystals demonstrate at hydrostatic compression insulator-to-metal (MIT) transition. Additionally, some representatives of this class of vdW crystals have been reported to display superconductivity \cite{Wang_2018}.

For Fe$_2$P$_2$S$_6$ and Fe$_2$P$_2$Se$_6$ crystals, the pressure-induced MIT is accompanied by volume collapse that is related to spin crossover from high-spin to low-spin state of Fe$^{2+}$ cations \cite{Wang_2018, Zheng_2019}. The unit cell reduction and the shift of phosphorus atoms into sulfurs plane inside of the PS$_3$ unit, together with appeared Fe-Fe distances nonequivalence leads to the broadening of both valence and conduction bands, resulting in the band gap collapse. Such behavior is accompanied by a significant contribution of phosphorus 3p-states at the bottom of the conduction band \cite{Wang_2018, Evarestov_2020}.

Also, for the Mn$_2$P$_2$S$_6$ compound, the first order isostructural phase transition accompanied by Mn-Mn dimerization together with 2D-3D crossover in the chemical bonding \cite{Wang_2018, PhysRevLett.123.236401} with volume collapse is crucial for the occurrence of metallization. Thus, the Mn$_2$P$_2$S$_6$ crystal is a clear example of structurally assisted bandwidth controlled MIT \cite{PhysRevLett.123.236401}.

In contrast, for the Ni$_2$P$_2$S$_6$ and Ni$_2$P$_2$Se$_6$ the MIT does not involve simultaneous structural changes, and these compounds are rare examples of the electronically driven bandwidth controlled transitions \cite{Wang_2018, Zheng_2019, PhysRevLett.123.236401}. 

The 3D Sn$_2$P$_2$S$_6$ molecular crystals exhibit ferroelectric structural transition with dipole ordering following the three-well local potential for the spontaneous polarization  \cite{PhysRevLett.99.207601}. 
Substitution of the Sn cation site by Pb \cite{PhysStatusSolidiB.253.384} or hydrostatic compression \cite{Ondrejkovic2012, Ondrejkovic2013} splits the temperature T$_0$ of the continuous transition at the tricritical point with P$_{TCP}$~=~0.6~GPa and T$_{TCP}$~$\sim$~250~K. At higher pressures the first order paraelectric-to-ferroelectric phase transition goes down till 0~K and the paraelectric ground state is stabilized for P~$>$~1.5~GPa \cite{Ondrejkovic2013}. The Pb$_2$P$_2$S$_6$ isostructural crystal with fully substituted Sn sites is incipient ferroelectric at normal pressure condition and here the quantum paraelectric state is observed at low temperatures \cite{Zamaraite2020}.

The high-pressure centrosymmetric ground state in Sn$_2$P$_2$S$_6$ is caused by the suppression of the second order Jahn-Teller effect, which is related to the stereochemical activity of the 5s$^2$ electron lone pair localized on Sn$^{2+}$ cations   \cite{PhysRevLett.99.207601}, together with the elimination of the charge disproportionation (P$^{4+}$+P$^{4+}$) $-$ (P$^{3+}$+P$^{5+}$) on the phosphorus P-P dimer sites inside anionic molecular units \cite{Zamaraite2020}. With further compression of the Sn$_2$P$_2$S$_6$ crystal, its bandgap E$_{g}$ decreases, and the electrical resistivity shows metallic behavior above 38~GPa \cite{shchennikov2011, ovsyannikov2013}. For the selenide compound Sn$_2$P$_2$Se$_6$, the metallization occurs at lower pressures above 20~GPa, while the Pb$_2$P$_2$S$_6$ crystal is still semiconductive at 50~GPa \cite{ovsyannikov2017}.

The Raman spectroscopy studies \cite{ovsyannikov2013,  ovsyannikov2017} reveal the clear signatures of structural changes in Sn$_2$P$_2$S$_6$ with increased hydrostatic pressure. The pressure-induced structural changes can be traced in the evolution of the Raman spectral line that is placed near 378~cm$^{-1}$ at ambient conditions. This line is related to the internal vibrations of P$_2$S$_6$ groups caused by the stretching of the P-P bond merging two PS$_3$ structural pyramids. The disappearance of this spectral line can be related to the rearrangement of the quasi-molecular P$_2$S$_6$ anionic structural groups and the appearance of the crystal lattice with a new topology.

The Sn$_2$P$_2$S$_6$ crystal has recently been reported \cite{yue2021} to exhibit superconductivity in its metallic state, with a transition temperature of T$_{C}$ $\approx$ 2.2 K at 31.7 GPa. This temperature slightly increases to 2.8 K when the pressure reaches 48.9 GPa \cite{yue2021}.

Obviously, the metallization in Sn$_2$P$_2$S$_6$ crystals belongs to the case of structurally assisted bandwidth controlled MIT. While the appearance of $-$P-P$-$ chains with increased P-P bonds length inside of the P$_2$S$_6$ molecular unit strongly rearranges the electron density of states (DOS) near the Fermi level in Fe$_2$P$_2$S$_6$ crystals \cite{Evarestov_2020}, breaking of P-P dimers can also effectively broaden the bonding and antibonding orbitals of P$_2$S$_6$ groups, resulting in higher density of states near the Fermi level \cite{jia2011, imai2014}, which can also favor the superconductivity \cite{chen2022}.

\section{Computational and Experimental methods}

With aim to establish the structural evolution of the Sn$_2$P$_2$S$_6$ crystal at compression and during transition from the semiconducting to the metallic state, we performed high-pressure X-ray diffraction experiments on the PETRA III P02.2 beamline at DESY, Hamburg. Monocrystalline samples were loaded into diamond anvil cell. The pressure was varied in the range from ambient to 55~GPa. The experimentally determined pressure dependence of Sn$_2$P$_2$S$_6$ structural characteristics is compared with results of density functional theory (DFT) calculations. The obtained experimental and theoretical results on structural changes at the MIT in Sn$_2$P$_2$S$_6$ are compared with previously published data \cite{yue2021} on pressure-induced superconductivity in this compound.

The unit cell of the Sn$_2$P$_2$S$_6$ consists of two formula units, i.e. 20 atoms. To predict possible high-pressure structures, we applied an evolutionary algorithm as implemented in the \textsc{USPEX} code \cite{uspex_1,uspex_2,uspex_3} in combination with DFT calculations using the \textsc{VASP} code \cite{VASP_Kresse:1993,VASP_Kresse:1996}. In order to get the most detailed sampling of possible structural configurations, we used several steps with different initial assumptions: (i) to preserve the local atomic configuration to be close as most as possible to its initial ambient structure we consider the Sn$_2$P$_2$S$_6$ as molecular structure with separated Sn$^{2+}$ cations and P$_2$S$_6^{4-}$ anions; (ii) we assume breaking of P-P bonds in P$_2$P$_6$ molecular unit resulting in two PS$_3$ molecular pyramids; (iii) we ensured that all atoms behaved independently to allow for the most complex rearrangements leading to the ground state structure; (iv) in order to address only kinetically related reconstructions we applied evolutionary metadynamics algorithm. For the cases (i)-(iii) we consider stoichiometries ranging from 5 to 160 atoms (sampling possibilities with the doubled in three directions unit cells) with all intermediate steps according SnPS$_3$ molecular unit. In evolutionary metadynamics we consider cases with one unit cell (i.e. $\Gamma$-point phonon distortion) and $2\times2\times2$ supercell, which allows distortions also for phonons on the Brillouin zone wedges.

All \textit{ab initio} results presented below were obtained within the local-density approximation (LDA) \cite{LDA} to account the exchange and correlation interactions. We have also checked the results with SCAN potential \cite{SCAN}.
The structural parameters of the candidate off-springs were optimized with \textsc{VASP} code using a plane-wave cutoff of 350~eV. The Brillouin zone was sampled with uniform $\Gamma$-centered k-point meshes with a resolution of $2 \pi \times 0.05$~\AA$^{-1}$. An energy convergence threshold was set to 10$^{^{-6}}$~eV, Hellman-Feynman forces were converged to threshold of 10$^{^{-6}}$~eV/\AA. 
In the evolutionary algorithm we allowed heredity (to generate 20\% of the off-spring structures for the next generation), soft-mutation (50\%), lattice-mutation (20\%) and random structure (10\%) operators. 
For evolutionary metadynamics in the $2\times  2\times 2$ supercell we used a Gaussian with the width of 0.8~\AA\ and a height of 3800~\AA$^{3}$ $\cdot$ kbar, as well as 2~\AA\ for the soft-mode amplitudes.

\section{Results and Discussion}

\begin{figure*}
	\includegraphics{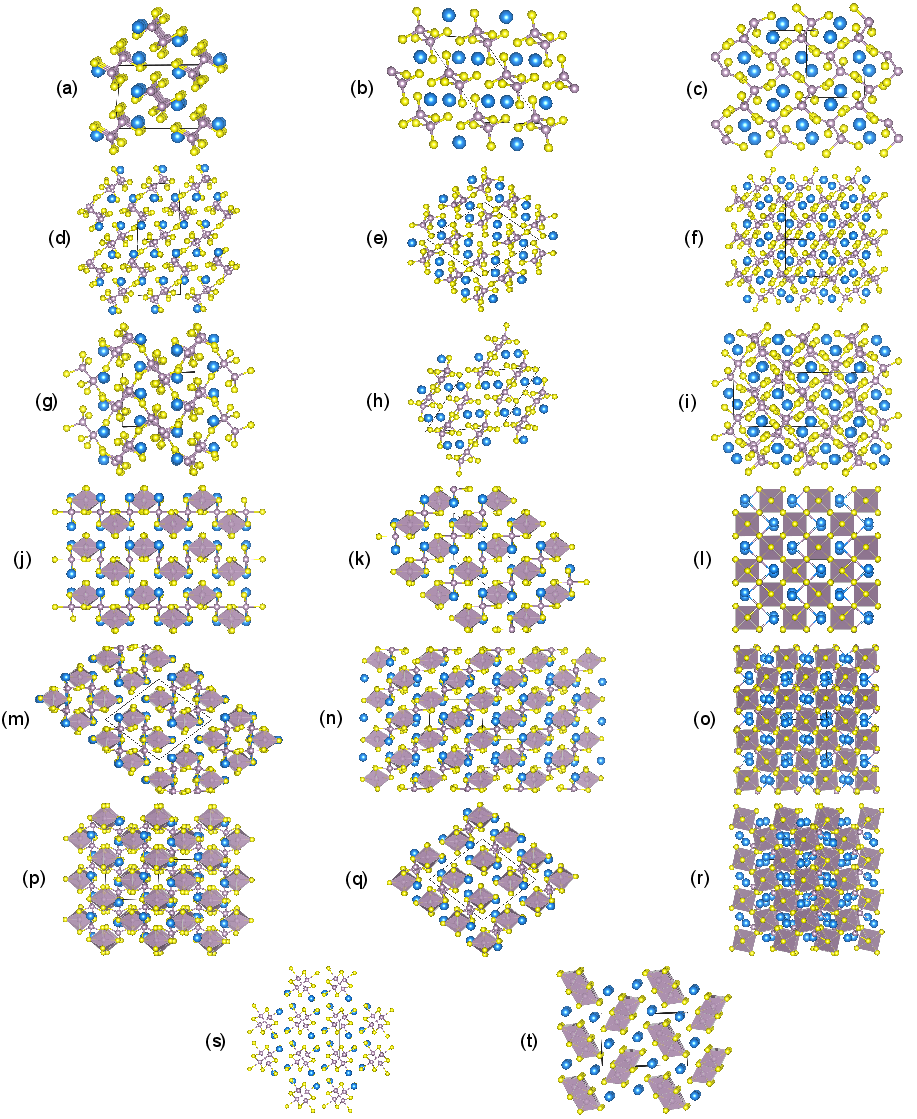}	
	\caption{\label{fig:structures}(Color online) 
		High-pressure candidates of crystalline structure of Sn$_2$P$_2$S$_6$ at $\sim$ 43~GPa:
		(a-c): Three projections of the ambient structure along orthogonal x, y, z axes with non-interacting P$_2$S$_6$ molecules. The unit cell is marked by a thin black line. Tin atoms are blue, sulfur atoms are yellow, and phosphorus atoms are light purple;
		(d-f): the same as in (a-c), but P$_2$S$_6$ molecules connected into chains;
		(g-i): the same as in previous cases, but with interacting chains of P$_2$S$_6$ molecules;
		(j-l): three projections of kinetically connected low-energy structure with space symmetry P2/c; here and below: P atoms in the octahedral sulfur environment are represented by light purple polygons.
		(m-o): three projections of kinetically connected low-energy toroidal structure with space symmetry P-1; 
		(p-f): three projections of the low-energy structure with symmetry P2$_1$/c, originating from the high-energy structure with P$_2$S$_6$ molecular chains and rings;
		(s) The most compact low energy P4$_2$/m structure with preserved P-P bond in molecular units;
		(t): Ground-state Pnma structure with two P atoms in the octahedral sulfur environment.
	}
\end{figure*}

\begin{figure}
	\includegraphics[width=0.94\hsize]{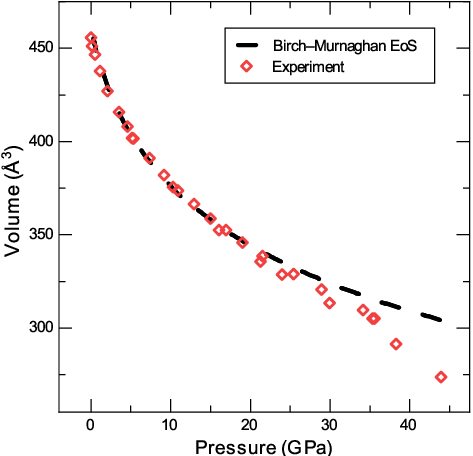}
	\caption{\label{fig:experiment}(Color online) Pressure dependence of the monoclinic unit cell volume of Sn$_2$P$_2$S$_6$ according to X-ray diffraction data and its fitting \cite{conference2019} using the Birch-Murnaghan equation of state.
	}
\end{figure}

\begin{figure}
	\includegraphics[width=0.93\hsize]{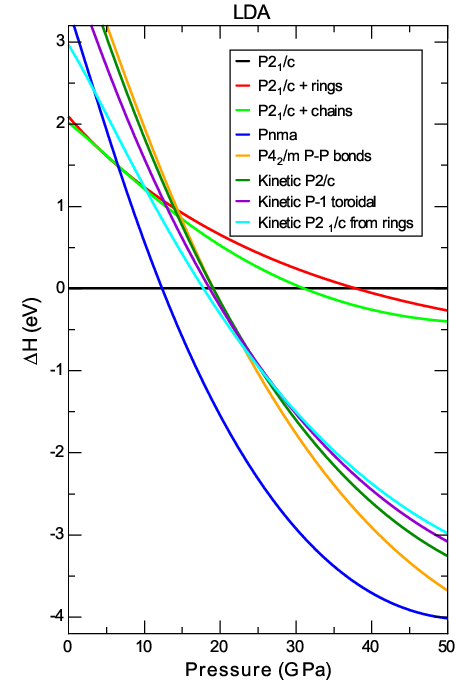}
	\caption{\label{fig:enthalpy}(Color online) Enthalpy difference for different phases of Sn$_2$P$_2$S$_6$ relative to ambient condition P2$_1$/c paraelectric phase.
	}
\end{figure}

Similarly to organic chemistry issue, where the structure of the product strongly depends on the structure of the reactant and the experimental conditions, defining the real reaction path from the vast set of possible combinations, the pressure-induced structural transformation in molecular Sn$_2$P$_2$S$_6$ is driven by kinetically compatible transformations. The pressure evolution of Raman spectra, reported in \cite{ovsyannikov2013} revealed five areas, where changes in Gr\"{u}neisen parameters of selected modes could be associated with the local structural transformations. As mentioned above, the mode $\sim$ 375~cm$^{-1}$ is a signature mode for the presence of P-P bond in the molecular unit (see Figure~\ref{fig:structures}a). The Raman intensity of this mode is significantly depressed in the compression regime at the pressures above $\sim$ 18~GPa, however the mode remains visible up to the MIT pressure at 38~GPa. In the decompression regime, the mode recovers when the pressure reaches $\sim$ 14.3~GPa, revealing significant hysteresis in the structural transformation. Based on X-ray diffraction data \cite{conference2019}, the unit cell volume of Sn$_2$P$_2$S$_6$ correlates well with the Birch-Murnaghan equation of state from ambient pressure to 20 GPa. However, at higher pressures, there is a clear deviation of the experimental data from the extrapolated curve of the equation of state (see Figure~\ref{fig:experiment}). In the same pressure range, a twinning of the diffraction reflexes was observed. This may indicate that complex structural rearrangements of the Sn$_2$P$_2$S$_6$ crystalline compound were induced by pressure. Molecular dynamics simulations and simulated annealing calculations, performed in the $2\times  2\times 2$ supercell (160 atoms) at moderate temperatures of 300 and 400~K, indicated significant disorder of tin atoms around their equilibrium positions and binding of molecular units first to the zig-zag chains (see Figure~\ref{fig:structures}d) and then, at higher pressures, to the pattern with connected neighboring chains (Figure~\ref{fig:structures}g). This occurs without any change in the structure of the P$_2$S$_6$ molecule, i.e. the P-P bond is still preserved. These local bond reconnections lead to an isostructural doubling of the unit cell volume, which might explain the formation of the doublets in the Raman spectra of selected modes (see, for example, the mode at 600~cm$^{-1}$ in the region above 28~GPa (region V) in Figure~4 of Ref.~\cite{ovsyannikov2013}). Although the thermal enthalpy results show that the ring-pattern remains metastable with respect to the chain-like structure (compare red and green lines in Figure~\ref{fig:enthalpy}), finite-temperature calculations clearly revealed the formation of bonds between chains as the pressure is further increased. As a result, one should consider three possible starting structures to calculate kinetically favored structural changes: (i) ambient-like, where isolated molecules do not directly interact, see Figure~\ref{fig:structures}a-c; (ii) chained structure, where one sulfur atom of the molecule interacts with the P atom of another molecule, forming a zig-zag pattern (Figure~\ref{fig:structures}d-f), and (iii) with rings pattern, where two sulfurs of one molecule interact with the P atoms of other molecules (Figure~\ref{fig:structures}g-i). Obviously, this bond, already formed in the initial structure, will define the transition path to the high-pressure, low-energy configuration. In fact, we found that the transition to the low-energy structures follows several common principles: P-P bond is broken in all scenarios; half of the P atoms occupy the crystallographic sites, octahedrally surrounded by S; the other half of the P atoms interact with both Sn and S atoms, and this interaction defines the differences in the resulting patterns.

We found out three possible phases, which differ in the position of this second half of the P atoms, located near the P's in sulfur octahedrons. The first pattern was obtained in evolutionary metadynamic simulations, when we considered a unit cell of ambient-like structure, quenched to pressures above 39~GPa. This pattern, with the transition path assigned only to $\Gamma$-point phonons, results in the P2/c structure (see Figure~\ref{fig:structures}j-l), and in LDA has the lowest energy among the considered kinetically connected phases (see Figure~\ref{fig:enthalpy}, dark green curve). However, taking the $2\times  2\times 2$ superstructure of the same quenched ambient-like phase as the initial one, the transition path according to the phonon modes from the Brillouin zone wedges favoring the breaking of P-P bonds with quadrupolar displacements of the second P atoms into the positions with two apical Sn atoms and four planar S atoms is revealed. This leads to the formation of the second pattern with interacting units, formerly belonging to four independent P$_2$S$_6$ molecules, resulting in the non-interacting rings (see Figure~\ref{fig:structures}m). These rings can be seen as large molecules with toroidal structure, forming a bulk molecular crystal structure. Therefore, the transition according to this transformation path can be described as pressure-induced toroidal ordering, which may have interesting topological properties.

\begin{figure}
	\includegraphics[width=0.93\hsize]{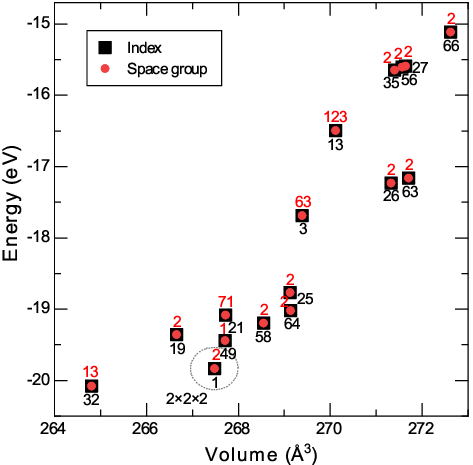}
	\caption{\label{fig:diagram}(Color online) Phase diagram calculated at pressure 43~GPa in combinatorial approach, for the structures with half P atoms in octahedral S environment, and the second half distributed in empty spaces with four planar S and two Sn apical sites in ambient-like unit cell with 20 atoms.
	}
\end{figure}

In the initial ambient-like structure, the individual P$_2$S$_6$ molecules show symmetry. This leads to uncertainty in the displacements of two P atoms, forming the P-P bond: which one will occupy the octahedral position in the S environment, and which one will go to the sites with two apical Sn atoms. To sample the possible occupancies, we compute the unit cell structure in the final form, keeping the octahedral coordinated position of P similar to P2/c configuration, but varying the occupancy of 12 empty spaces, filling them with two P atoms. This leads us to 66 combinations (without taking symmetries into account), with the energy, presented in Figure~\ref{fig:diagram}. The figure also includes the energy for the P-1 toroidal phase, obtained from the supercell. Unfortunately, similar calculations for the most favorable pattern in the $2\times  2\times 2$ are impossible due to the huge number (over 6.6$\times$10$^{17}$) of configurations.

The third pattern originates from chain- and ring-like initial structures, and differs from the previously described ones. Similar to the previous cases, half of the P atoms occupy octahedral S-coordinated sites. The pattern formation of this phosphorus is common for all three cases. However, in the last case, the second P atoms from connected molecules (see Figures~\ref{fig:structures}e and h) are pushed towards the connections. The difference is most obvious when Figure~\ref{fig:structures}q is compared with (k) and (n) of previous patterns: whereas in the case of P2/c and P-1 patterns the P atoms in this projection form infinite planes following the repetitive scheme "octahedral"-"planar" positions, in the last case the scheme is "octahedral"-"planar"-"planar"-"octahedral" position, with broken periodicity. In fact, the "planar" position, which in the previous cases is symmetric with respect to two apical Sn atoms, is broken in the third pattern, and the P atoms interact with four S atoms, but with only one Sn (see Figure~\ref{fig:structures}r).

One should note, that the evolutionary algorithm also revealed the low-energy configuration, where P-P bonds are still conserved (see Figure~\ref{fig:structures}s). Molecular units in this structure form cross-patterns, which seems to be kinetically forbidden and is not realized at ambient temperature conditions. This configuration is the most symmetric structure among all low-energy candidates obtained, and it is, surprisingly, the most compact. However, the simulated X-ray pattern differs significantly from the experimental results, similar to the ground-state Pnma structure with two P atoms in the octahedral sulfur environment (see Figure~\ref{fig:structures}t), which excludes the possible scenario of its formation, and indicates that the kinetic consideration of the possible transition is more realistic for the experimental condition performed at room temperature. However, experiments at elevated temperatures might reveal the structural transformation to this high-pressure ground state.

Calculations, performed using SCAN functional, revealed qualitatively different scenario: the ambient-like phase remain stable over chain- and ring-phases in experimental range of pressures, and the toroidal phase is the lowest-energy one among considered kinetically-driven phases. Therefore, from the theoretical point of view, it is difficult to judge the resulting phase. Moreover, as it is seen in Figure~\ref{fig:diagram}, a lot of possible patterns occupies narrow ($\sim$ 0.1~eV per atom) energy range above the lowest energy P2/c configuration, leading to the conclusion, that the resulting phase might be a combination of possible patterns, which could be seen in the experiment as an average disordered phase. The positions of Sn atoms are also scattered around high-symmetry sites in all considered phases (compare Sn position in Figures~\ref{fig:structures}l, o, r with the ambient phase (c)) already in the ground state calculated at 0~K. Also, for all low temperature phases molecular dynamics simulation revealed high displacements amplitude of Sn atoms already at room temperature. All together, one should expect to see an averaged structure with highly occupied octahedral positions of P, and partially occupied "planar" positions, however, the partial occupancy might be significantly different, indicating preferential sites for the second P according to theoretically reported patterns.

All considered low-energy high-pressure phases are predicted to be metallic. However, the toroidal phase is of particular interest due to its topological properties, which we will illustrate in the calculated electronic properties to highlight the potential of this class of materials to be topological, which might be related to the observed superconductivity in high-pressure metallic phase \cite{yue2021}.

\section{Conclusions}
The MM$'$P$_2$X$_6$ compounds feature molecular anions [P$_2$S(Se)$_6$]$^{4-}$ arranged into periodic lattices of varying dimensionality at ambient pressure. These compounds exhibit different types of ferroicity and multiferroicity that interact with insulator (semiconductor) to metal state transitions at room temperature. Near the boundary between insulating and metallic phases at low temperatures, a superconducting state emerges. Compressing M$_2$P$_2$X$_6$ layered compounds alters the dimensionality of the crystal lattice by narrowing the van der Waals gap and creating covalent chemical bonding between atoms from the nearest structural layer. The 2D to 3D crossover is accompanied by the sliding together of the nearest structural layers and the reconstruction of the space arrangement of the layer-built atoms. This "three-levels" atomic space reconfiguration reflects important variations in the atomic valence orbitals hybridization, which modifies the electronic densities near the top of the valence band and at the bottom of the conductivity band, resulting in the collapse of the energy gap.

Based on our investigations, the Sn$_2$P$_2$S$_6$ compound with two sublattices of (P$_2$S$_6$)$^{4-}$ anionic molecules arranged in a 3D structure, presents a wide range of possibilities for structural rearrangement under compression. It is noteworthy that P$_2$S$_6$ molecules can transform into covalently bonded PS$_6$ and Sn$_2$PS$_4$ structural groups, which is related to the rehybridization of valence orbitals. The structural instability of P$_2$S$_6$ anionic molecules accompanies the pressure-induced metallization of the Sn$_2$P$_2$S$_6$ compound. This provides clear evidence of strong electron-phonon coupling involved in the rearrangement of the electronic structure. As in the case of ferroelectric ordering with cooling at ambient pressure, which is assisted by nonlinear coupling between the fully symmetrical internal vibrations of (P$_2$S$_6$)$^{4-}$ anions with relative shifts of Sn$^{2+}$ cations \cite{PhysRevLett.99.207601}, the insulator-to-metal transition is also related to the instability of the structural arrangement of P$_2$S$_6$ molecules. 

Increasing the flexibility of P$_2$S$_6$ molecules and the polarizability of the crystal lattice when approaching metallization pressure can create conditions for the appearance of electronic pairs and the presence of superconductivity, as predicted by models in the strong coupling regime \cite{chakraverty1979, PhysRevB.19.3593, micnas:jpa-00229287, PhysRevLett.61.2713, Mott1993, Rowley_2014, Volkov_2022}.

\begin{acknowledgments}
K.Z.R. gratefully acknowledge the financial support by Deutsche Forschungsgemeinschaft (DFG) through SFB 917 ``Nanoswitches'', as well as the computing time granted through JARA on the supercomputer JURECA \cite{jureca} at Forschungszentrum J\"ulich and JARA-HPC Partition (project jara0126). Visualisation was made with the help of {\textsc VESTA} code \cite{VESTA}. {\textsc FINDSYM} package \cite{FINDSYM} was used for symmetry analysis of obtained structures. M.M. and Yu.M.V. acknowledge staffs of the PETRA III P02.2 beamline at DESY for providing the high pressure XRD experiments and the project CALIPSOplus under the Grant Agreement 730872 from the EU Framework Program for Research and Innovation HORIZON 2020. We thank Dr.~Ivetta Slipukhina for the critical reading of the manuscript and valuable suggestions. 
\end{acknowledgments}

\bibliographystyle{apsrev4-2}
\bibliography{references.bib}

\begin{thebibliography}{45}%
\makeatletter
\providecommand \@ifxundefined [1]{%
 \@ifx{#1\undefined}
}%
\providecommand \@ifnum [1]{%
 \ifnum #1\expandafter \@firstoftwo
 \else \expandafter \@secondoftwo
 \fi
}%
\providecommand \@ifx [1]{%
 \ifx #1\expandafter \@firstoftwo
 \else \expandafter \@secondoftwo
 \fi
}%
\providecommand \natexlab [1]{#1}%
\providecommand \enquote  [1]{``#1''}%
\providecommand \bibnamefont  [1]{#1}%
\providecommand \bibfnamefont [1]{#1}%
\providecommand \citenamefont [1]{#1}%
\providecommand \href@noop [0]{\@secondoftwo}%
\providecommand \href [0]{\begingroup \@sanitize@url \@href}%
\providecommand \@href[1]{\@@startlink{#1}\@@href}%
\providecommand \@@href[1]{\endgroup#1\@@endlink}%
\providecommand \@sanitize@url [0]{\catcode `\\12\catcode `\$12\catcode
  `\&12\catcode `\#12\catcode `\^12\catcode `\_12\catcode `\%12\relax}%
\providecommand \@@startlink[1]{}%
\providecommand \@@endlink[0]{}%
\providecommand \url  [0]{\begingroup\@sanitize@url \@url }%
\providecommand \@url [1]{\endgroup\@href {#1}{\urlprefix }}%
\providecommand \urlprefix  [0]{URL }%
\providecommand \Eprint [0]{\href }%
\providecommand \doibase [0]{https://doi.org/}%
\providecommand \selectlanguage [0]{\@gobble}%
\providecommand \bibinfo  [0]{\@secondoftwo}%
\providecommand \bibfield  [0]{\@secondoftwo}%
\providecommand \translation [1]{[#1]}%
\providecommand \BibitemOpen [0]{}%
\providecommand \bibitemStop [0]{}%
\providecommand \bibitemNoStop [0]{.\EOS\space}%
\providecommand \EOS [0]{\spacefactor3000\relax}%
\providecommand \BibitemShut  [1]{\csname bibitem#1\endcsname}%
\let\auto@bib@innerbib\@empty
\bibitem [{\citenamefont {Brec}(1986)}]{brec1986review}%
  \BibitemOpen
  \bibfield  {author} {\bibinfo {author} {\bibfnamefont {R.}~\bibnamefont
  {Brec}},\ }\href
  {https://doi.org/https://doi.org/10.1016/0167-2738(86)90055-X} {\bibfield
  {journal} {\bibinfo  {journal} {Solid State Ionics}\ }\textbf {\bibinfo
  {volume} {22}},\ \bibinfo {pages} {3} (\bibinfo {year} {1986})}\BibitemShut
  {NoStop}%
\bibitem [{\citenamefont {Samal}\ \emph {et~al.}(2021)\citenamefont {Samal},
  \citenamefont {Sanyal}, \citenamefont {Chakraborty},\ and\ \citenamefont
  {Rout}}]{samal2021}%
  \BibitemOpen
  \bibfield  {author} {\bibinfo {author} {\bibfnamefont {R.}~\bibnamefont
  {Samal}}, \bibinfo {author} {\bibfnamefont {G.}~\bibnamefont {Sanyal}},
  \bibinfo {author} {\bibfnamefont {B.}~\bibnamefont {Chakraborty}},\ and\
  \bibinfo {author} {\bibfnamefont {C.~S.}\ \bibnamefont {Rout}},\ }\href
  {https://doi.org/10.1039/D0TA09752G} {\bibfield  {journal} {\bibinfo
  {journal} {J. Mater. Chem. A}\ }\textbf {\bibinfo {volume} {9}},\ \bibinfo
  {pages} {2560} (\bibinfo {year} {2021})}\BibitemShut {NoStop}%
\bibitem [{\citenamefont {Susner}\ \emph {et~al.}(2017)\citenamefont {Susner},
  \citenamefont {Chyasnavichyus}, \citenamefont {McGuire}, \citenamefont
  {Ganesh},\ and\ \citenamefont {Maksymovych}}]{susner2017}%
  \BibitemOpen
  \bibfield  {author} {\bibinfo {author} {\bibfnamefont {M.~A.}\ \bibnamefont
  {Susner}}, \bibinfo {author} {\bibfnamefont {M.}~\bibnamefont
  {Chyasnavichyus}}, \bibinfo {author} {\bibfnamefont {M.~A.}\ \bibnamefont
  {McGuire}}, \bibinfo {author} {\bibfnamefont {P.}~\bibnamefont {Ganesh}},\
  and\ \bibinfo {author} {\bibfnamefont {P.}~\bibnamefont {Maksymovych}},\
  }\href {https://doi.org/https://doi.org/10.1002/adma.201602852} {\bibfield
  {journal} {\bibinfo  {journal} {Advanced Materials}\ }\textbf {\bibinfo
  {volume} {29}},\ \bibinfo {pages} {1602852} (\bibinfo {year}
  {2017})}\BibitemShut {NoStop}%
\bibitem [{\citenamefont {Carpentier}\ and\ \citenamefont
  {Nitsche}(1974)}]{carpentier1974}%
  \BibitemOpen
  \bibfield  {author} {\bibinfo {author} {\bibfnamefont {C.}~\bibnamefont
  {Carpentier}}\ and\ \bibinfo {author} {\bibfnamefont {R.}~\bibnamefont
  {Nitsche}},\ }\href
  {https://doi.org/https://doi.org/10.1016/0025-5408(74)90207-4} {\bibfield
  {journal} {\bibinfo  {journal} {Materials Research Bulletin}\ }\textbf
  {\bibinfo {volume} {9}},\ \bibinfo {pages} {401} (\bibinfo {year}
  {1974})}\BibitemShut {NoStop}%
\bibitem [{\citenamefont {Liubachko}\ \emph {et~al.}(2020)\citenamefont
  {Liubachko}, \citenamefont {Oleaga}, \citenamefont {Salazar}, \citenamefont
  {Yevych}, \citenamefont {Kohutych},\ and\ \citenamefont
  {Vysochanskii}}]{liubachko2020}%
  \BibitemOpen
  \bibfield  {author} {\bibinfo {author} {\bibfnamefont {V.}~\bibnamefont
  {Liubachko}}, \bibinfo {author} {\bibfnamefont {A.}~\bibnamefont {Oleaga}},
  \bibinfo {author} {\bibfnamefont {A.}~\bibnamefont {Salazar}}, \bibinfo
  {author} {\bibfnamefont {R.}~\bibnamefont {Yevych}}, \bibinfo {author}
  {\bibfnamefont {A.}~\bibnamefont {Kohutych}},\ and\ \bibinfo {author}
  {\bibfnamefont {Y.}~\bibnamefont {Vysochanskii}},\ }\href
  {https://doi.org/10.1103/PhysRevB.101.224110} {\bibfield  {journal} {\bibinfo
   {journal} {Phys. Rev. B}\ }\textbf {\bibinfo {volume} {101}},\ \bibinfo
  {pages} {224110} (\bibinfo {year} {2020})}\BibitemShut {NoStop}%
\bibitem [{\citenamefont {Maisonneuve}\ \emph {et~al.}(1997)\citenamefont
  {Maisonneuve}, \citenamefont {Cajipe}, \citenamefont {Simon}, \citenamefont
  {Von Der~Muhll},\ and\ \citenamefont {Ravez}}]{maisonneuve1997}%
  \BibitemOpen
  \bibfield  {author} {\bibinfo {author} {\bibfnamefont {V.}~\bibnamefont
  {Maisonneuve}}, \bibinfo {author} {\bibfnamefont {V.~B.}\ \bibnamefont
  {Cajipe}}, \bibinfo {author} {\bibfnamefont {A.}~\bibnamefont {Simon}},
  \bibinfo {author} {\bibfnamefont {R.}~\bibnamefont {Von Der~Muhll}},\ and\
  \bibinfo {author} {\bibfnamefont {J.}~\bibnamefont {Ravez}},\ }\href
  {https://doi.org/10.1103/PhysRevB.56.10860} {\bibfield  {journal} {\bibinfo
  {journal} {Phys. Rev. B}\ }\textbf {\bibinfo {volume} {56}},\ \bibinfo
  {pages} {10860} (\bibinfo {year} {1997})}\BibitemShut {NoStop}%
\bibitem [{\citenamefont {Dziaugys}\ \emph {et~al.}(2020)\citenamefont
  {Dziaugys}, \citenamefont {Kelley}, \citenamefont {Brehm}, \citenamefont
  {Tao}, \citenamefont {Puretzky}, \citenamefont {Feng}, \citenamefont
  {O'Hara}, \citenamefont {Neumayer}, \citenamefont {Chyasnavichyus},
  \citenamefont {Eliseev}, \citenamefont {Banys}, \citenamefont {Vysochanskii},
  \citenamefont {Ye}, \citenamefont {Chakoumakos}, \citenamefont {McGuire},
  \citenamefont {Kalinin}, \citenamefont {Ganesh}, \citenamefont {Balke},
  \citenamefont {Pantelides}, \citenamefont {Morozovska},\ and\ \citenamefont
  {Maksymovych}}]{dziaugys2020}%
  \BibitemOpen
  \bibfield  {author} {\bibinfo {author} {\bibfnamefont {A.}~\bibnamefont
  {Dziaugys}}, \bibinfo {author} {\bibfnamefont {K.}~\bibnamefont {Kelley}},
  \bibinfo {author} {\bibfnamefont {J.~A.}\ \bibnamefont {Brehm}}, \bibinfo
  {author} {\bibfnamefont {L.}~\bibnamefont {Tao}}, \bibinfo {author}
  {\bibfnamefont {A.}~\bibnamefont {Puretzky}}, \bibinfo {author}
  {\bibfnamefont {T.}~\bibnamefont {Feng}}, \bibinfo {author} {\bibfnamefont
  {A.}~\bibnamefont {O'Hara}}, \bibinfo {author} {\bibfnamefont
  {S.}~\bibnamefont {Neumayer}}, \bibinfo {author} {\bibfnamefont
  {M.}~\bibnamefont {Chyasnavichyus}}, \bibinfo {author} {\bibfnamefont
  {E.~A.}\ \bibnamefont {Eliseev}}, \bibinfo {author} {\bibfnamefont
  {J.}~\bibnamefont {Banys}}, \bibinfo {author} {\bibfnamefont
  {Y.}~\bibnamefont {Vysochanskii}}, \bibinfo {author} {\bibfnamefont
  {F.}~\bibnamefont {Ye}}, \bibinfo {author} {\bibfnamefont {B.~C.}\
  \bibnamefont {Chakoumakos}}, \bibinfo {author} {\bibfnamefont {M.~A.}\
  \bibnamefont {McGuire}}, \bibinfo {author} {\bibfnamefont {S.~V.}\
  \bibnamefont {Kalinin}}, \bibinfo {author} {\bibfnamefont {P.}~\bibnamefont
  {Ganesh}}, \bibinfo {author} {\bibfnamefont {N.}~\bibnamefont {Balke}},
  \bibinfo {author} {\bibfnamefont {S.~T.}\ \bibnamefont {Pantelides}},
  \bibinfo {author} {\bibfnamefont {A.~N.}\ \bibnamefont {Morozovska}},\ and\
  \bibinfo {author} {\bibfnamefont {P.}~\bibnamefont {Maksymovych}},\ }\href
  {https://doi.org/10.1038/s41467-020-17137-0} {\bibfield  {journal} {\bibinfo
  {journal} {Nature Communications}\ }\textbf {\bibinfo {volume} {11}},\
  \bibinfo {pages} {3623} (\bibinfo {year} {2020})}\BibitemShut {NoStop}%
\bibitem [{\citenamefont {Ouvrard}\ \emph {et~al.}(1985)\citenamefont
  {Ouvrard}, \citenamefont {Brec},\ and\ \citenamefont {Rouxel}}]{ouvrard1985}%
  \BibitemOpen
  \bibfield  {author} {\bibinfo {author} {\bibfnamefont {G.}~\bibnamefont
  {Ouvrard}}, \bibinfo {author} {\bibfnamefont {R.}~\bibnamefont {Brec}},\ and\
  \bibinfo {author} {\bibfnamefont {J.}~\bibnamefont {Rouxel}},\ }\href
  {https://doi.org/10.1016/0025-5408(85)90092-3} {\bibfield  {journal}
  {\bibinfo  {journal} {Materials Research Bulletin}\ }\textbf {\bibinfo
  {volume} {20}},\ \bibinfo {pages} {1181} (\bibinfo {year}
  {1985})}\BibitemShut {NoStop}%
\bibitem [{\citenamefont {Macutkevic}\ \emph {et~al.}(2008)\citenamefont
  {Macutkevic}, \citenamefont {Banys}, \citenamefont {Grigalaitis},\ and\
  \citenamefont {Vysochanskii}}]{macutkevic2008}%
  \BibitemOpen
  \bibfield  {author} {\bibinfo {author} {\bibfnamefont {J.}~\bibnamefont
  {Macutkevic}}, \bibinfo {author} {\bibfnamefont {J.}~\bibnamefont {Banys}},
  \bibinfo {author} {\bibfnamefont {R.}~\bibnamefont {Grigalaitis}},\ and\
  \bibinfo {author} {\bibfnamefont {Y.}~\bibnamefont {Vysochanskii}},\ }\href
  {https://doi.org/10.1103/PhysRevB.78.064101} {\bibfield  {journal} {\bibinfo
  {journal} {Phys. Rev. B}\ }\textbf {\bibinfo {volume} {78}},\ \bibinfo
  {pages} {064101} (\bibinfo {year} {2008})}\BibitemShut {NoStop}%
\bibitem [{\citenamefont {Kleemann}\ \emph {et~al.}(2011)\citenamefont
  {Kleemann}, \citenamefont {Shvartsman}, \citenamefont {Borisov},
  \citenamefont {Banys},\ and\ \citenamefont {Vysochanskii}}]{kleemann2011}%
  \BibitemOpen
  \bibfield  {author} {\bibinfo {author} {\bibfnamefont {W.}~\bibnamefont
  {Kleemann}}, \bibinfo {author} {\bibfnamefont {V.~V.}\ \bibnamefont
  {Shvartsman}}, \bibinfo {author} {\bibfnamefont {P.}~\bibnamefont {Borisov}},
  \bibinfo {author} {\bibfnamefont {J.}~\bibnamefont {Banys}},\ and\ \bibinfo
  {author} {\bibfnamefont {Y.~M.}\ \bibnamefont {Vysochanskii}},\ }\href
  {https://doi.org/10.1103/PhysRevB.84.094411} {\bibfield  {journal} {\bibinfo
  {journal} {Phys. Rev. B}\ }\textbf {\bibinfo {volume} {84}},\ \bibinfo
  {pages} {094411} (\bibinfo {year} {2011})}\BibitemShut {NoStop}%
\bibitem [{\citenamefont {Peschanskii}\ \emph {et~al.}(2019)\citenamefont
  {Peschanskii}, \citenamefont {Babuka}, \citenamefont {Glukhov}, \citenamefont
  {Makowska-Janusik}, \citenamefont {Gnatchenko},\ and\ \citenamefont
  {Vysochanskii}}]{peschanskii2019}%
  \BibitemOpen
  \bibfield  {author} {\bibinfo {author} {\bibfnamefont {A.~V.}\ \bibnamefont
  {Peschanskii}}, \bibinfo {author} {\bibfnamefont {T.~Y.}\ \bibnamefont
  {Babuka}}, \bibinfo {author} {\bibfnamefont {K.~E.}\ \bibnamefont {Glukhov}},
  \bibinfo {author} {\bibfnamefont {M.}~\bibnamefont {Makowska-Janusik}},
  \bibinfo {author} {\bibfnamefont {S.~L.}\ \bibnamefont {Gnatchenko}},\ and\
  \bibinfo {author} {\bibfnamefont {Y.~M.}\ \bibnamefont {Vysochanskii}},\
  }\href {https://doi.org/https://doi.org/10.1063/1.5125909} {\bibfield
  {journal} {\bibinfo  {journal} {Low Temperature Physics}\ }\textbf {\bibinfo
  {volume} {45}},\ \bibinfo {pages} {1082} (\bibinfo {year}
  {2019})}\BibitemShut {NoStop}%
\bibitem [{\citenamefont {Wang}\ \emph {et~al.}(2018)\citenamefont {Wang},
  \citenamefont {Ying}, \citenamefont {Zhou}, \citenamefont {Sun},
  \citenamefont {Wen}, \citenamefont {Zhou}, \citenamefont {Li}, \citenamefont
  {Zhang}, \citenamefont {Han}, \citenamefont {Xiao}, \citenamefont {Chow},
  \citenamefont {Yang}, \citenamefont {Struzhkin}, \citenamefont {Zhao},\ and\
  \citenamefont {kwang Mao}}]{Wang_2018}%
  \BibitemOpen
  \bibfield  {author} {\bibinfo {author} {\bibfnamefont {Y.}~\bibnamefont
  {Wang}}, \bibinfo {author} {\bibfnamefont {J.}~\bibnamefont {Ying}}, \bibinfo
  {author} {\bibfnamefont {Z.}~\bibnamefont {Zhou}}, \bibinfo {author}
  {\bibfnamefont {J.}~\bibnamefont {Sun}}, \bibinfo {author} {\bibfnamefont
  {T.}~\bibnamefont {Wen}}, \bibinfo {author} {\bibfnamefont {Y.}~\bibnamefont
  {Zhou}}, \bibinfo {author} {\bibfnamefont {N.}~\bibnamefont {Li}}, \bibinfo
  {author} {\bibfnamefont {Q.}~\bibnamefont {Zhang}}, \bibinfo {author}
  {\bibfnamefont {F.}~\bibnamefont {Han}}, \bibinfo {author} {\bibfnamefont
  {Y.}~\bibnamefont {Xiao}}, \bibinfo {author} {\bibfnamefont {P.}~\bibnamefont
  {Chow}}, \bibinfo {author} {\bibfnamefont {W.}~\bibnamefont {Yang}}, \bibinfo
  {author} {\bibfnamefont {V.~V.}\ \bibnamefont {Struzhkin}}, \bibinfo {author}
  {\bibfnamefont {Y.}~\bibnamefont {Zhao}},\ and\ \bibinfo {author}
  {\bibfnamefont {H.}~\bibnamefont {kwang Mao}},\ }\href
  {https://doi.org/10.1038/s41467-018-04374-4} {\bibfield  {journal} {\bibinfo
  {journal} {Nature Communications}\ }\textbf {\bibinfo {volume} {9}},\
  \bibinfo {pages} {1914} (\bibinfo {year} {2018})}\BibitemShut {NoStop}%
\bibitem [{\citenamefont {Zheng}\ \emph {et~al.}(2019)\citenamefont {Zheng},
  \citenamefont {xing Jiang}, \citenamefont {xiong Xue}, \citenamefont {Dai},\
  and\ \citenamefont {Feng}}]{Zheng_2019}%
  \BibitemOpen
  \bibfield  {author} {\bibinfo {author} {\bibfnamefont {Y.}~\bibnamefont
  {Zheng}}, \bibinfo {author} {\bibfnamefont {X.}~\bibnamefont {xing Jiang}},
  \bibinfo {author} {\bibfnamefont {X.}~\bibnamefont {xiong Xue}}, \bibinfo
  {author} {\bibfnamefont {J.}~\bibnamefont {Dai}},\ and\ \bibinfo {author}
  {\bibfnamefont {Y.}~\bibnamefont {Feng}},\ }\href
  {https://doi.org/10.1103/physrevb.100.174102} {\bibfield  {journal} {\bibinfo
   {journal} {Physical Review B}\ }\textbf {\bibinfo {volume} {100}},\ \bibinfo
  {pages} {174102} (\bibinfo {year} {2019})}\BibitemShut {NoStop}%
\bibitem [{\citenamefont {Evarestov}\ and\ \citenamefont
  {Kuzmin}(2020)}]{Evarestov_2020}%
  \BibitemOpen
  \bibfield  {author} {\bibinfo {author} {\bibfnamefont {R.~A.}\ \bibnamefont
  {Evarestov}}\ and\ \bibinfo {author} {\bibfnamefont {A.}~\bibnamefont
  {Kuzmin}},\ }\href {https://doi.org/https://doi.org/10.1002/jcc.26178}
  {\bibfield  {journal} {\bibinfo  {journal} {Journal of Computational
  Chemistry}\ }\textbf {\bibinfo {volume} {41}},\ \bibinfo {pages} {1337}
  (\bibinfo {year} {2020})}\BibitemShut {NoStop}%
\bibitem [{\citenamefont {Kim}\ \emph {et~al.}(2019)\citenamefont {Kim},
  \citenamefont {Haule},\ and\ \citenamefont
  {Vanderbilt}}]{PhysRevLett.123.236401}%
  \BibitemOpen
  \bibfield  {author} {\bibinfo {author} {\bibfnamefont {H.-S.}\ \bibnamefont
  {Kim}}, \bibinfo {author} {\bibfnamefont {K.}~\bibnamefont {Haule}},\ and\
  \bibinfo {author} {\bibfnamefont {D.}~\bibnamefont {Vanderbilt}},\ }\href
  {https://doi.org/10.1103/PhysRevLett.123.236401} {\bibfield  {journal}
  {\bibinfo  {journal} {Phys. Rev. Lett.}\ }\textbf {\bibinfo {volume} {123}},\
  \bibinfo {pages} {236401} (\bibinfo {year} {2019})}\BibitemShut {NoStop}%
\bibitem [{\citenamefont {Rushchanskii}\ \emph {et~al.}(2007)\citenamefont
  {Rushchanskii}, \citenamefont {Vysochanskii},\ and\ \citenamefont
  {Strauch}}]{PhysRevLett.99.207601}%
  \BibitemOpen
  \bibfield  {author} {\bibinfo {author} {\bibfnamefont {K.~Z.}\ \bibnamefont
  {Rushchanskii}}, \bibinfo {author} {\bibfnamefont {Y.~M.}\ \bibnamefont
  {Vysochanskii}},\ and\ \bibinfo {author} {\bibfnamefont {D.}~\bibnamefont
  {Strauch}},\ }\href {https://doi.org/10.1103/PhysRevLett.99.207601}
  {\bibfield  {journal} {\bibinfo  {journal} {Phys. Rev. Lett.}\ }\textbf
  {\bibinfo {volume} {99}},\ \bibinfo {pages} {207601} (\bibinfo {year}
  {2007})}\BibitemShut {NoStop}%
\bibitem [{\citenamefont {Rushchanskii}\ \emph {et~al.}(2016)\citenamefont
  {Rushchanskii}, \citenamefont {Bilanych}, \citenamefont {Molnar},
  \citenamefont {Yevych}, \citenamefont {Kohutych}, \citenamefont
  {Perechinskii}, \citenamefont {Samulionis}, \citenamefont {Banys},\ and\
  \citenamefont {Vysochanskii}}]{PhysStatusSolidiB.253.384}%
  \BibitemOpen
  \bibfield  {author} {\bibinfo {author} {\bibfnamefont {K.~Z.}\ \bibnamefont
  {Rushchanskii}}, \bibinfo {author} {\bibfnamefont {R.~M.}\ \bibnamefont
  {Bilanych}}, \bibinfo {author} {\bibfnamefont {A.~A.}\ \bibnamefont
  {Molnar}}, \bibinfo {author} {\bibfnamefont {R.~M.}\ \bibnamefont {Yevych}},
  \bibinfo {author} {\bibfnamefont {A.~A.}\ \bibnamefont {Kohutych}}, \bibinfo
  {author} {\bibfnamefont {S.~I.}\ \bibnamefont {Perechinskii}}, \bibinfo
  {author} {\bibfnamefont {V.}~\bibnamefont {Samulionis}}, \bibinfo {author}
  {\bibfnamefont {J.}~\bibnamefont {Banys}},\ and\ \bibinfo {author}
  {\bibfnamefont {Y.~M.}\ \bibnamefont {Vysochanskii}},\ }\href
  {https://doi.org/10.1002/pssb.201552403} {\bibfield  {journal} {\bibinfo
  {journal} {Phys. Status Solidi B}\ }\textbf {\bibinfo {volume} {253}},\
  \bibinfo {pages} {384} (\bibinfo {year} {2016})}\BibitemShut {NoStop}%
\bibitem [{\citenamefont {Ondrejkovic}\ \emph {et~al.}(2012)\citenamefont
  {Ondrejkovic}, \citenamefont {Kempa}, \citenamefont {Vysochanskii},
  \citenamefont {Saint-Gr\'egoire}, \citenamefont {Bourges}, \citenamefont
  {Rushchanskii},\ and\ \citenamefont {Hlinka}}]{Ondrejkovic2012}%
  \BibitemOpen
  \bibfield  {author} {\bibinfo {author} {\bibfnamefont {P.}~\bibnamefont
  {Ondrejkovic}}, \bibinfo {author} {\bibfnamefont {M.}~\bibnamefont {Kempa}},
  \bibinfo {author} {\bibfnamefont {Y.}~\bibnamefont {Vysochanskii}}, \bibinfo
  {author} {\bibfnamefont {P.}~\bibnamefont {Saint-Gr\'egoire}}, \bibinfo
  {author} {\bibfnamefont {P.}~\bibnamefont {Bourges}}, \bibinfo {author}
  {\bibfnamefont {K.~Z.}\ \bibnamefont {Rushchanskii}},\ and\ \bibinfo {author}
  {\bibfnamefont {J.}~\bibnamefont {Hlinka}},\ }\href
  {https://doi.org/10.1103/PhysRevB.86.224106} {\bibfield  {journal} {\bibinfo
  {journal} {Phys. Rev. B}\ }\textbf {\bibinfo {volume} {86}},\ \bibinfo
  {pages} {224106} (\bibinfo {year} {2012})}\BibitemShut {NoStop}%
\bibitem [{\citenamefont {Ondrejkovic}\ \emph {et~al.}(2013)\citenamefont
  {Ondrejkovic}, \citenamefont {Guennou}, \citenamefont {Kempa}, \citenamefont
  {Vysochanskii}, \citenamefont {Garbarino},\ and\ \citenamefont
  {Hlinka}}]{Ondrejkovic2013}%
  \BibitemOpen
  \bibfield  {author} {\bibinfo {author} {\bibfnamefont {P.}~\bibnamefont
  {Ondrejkovic}}, \bibinfo {author} {\bibfnamefont {M.}~\bibnamefont
  {Guennou}}, \bibinfo {author} {\bibfnamefont {M.}~\bibnamefont {Kempa}},
  \bibinfo {author} {\bibfnamefont {Y.}~\bibnamefont {Vysochanskii}}, \bibinfo
  {author} {\bibfnamefont {G.}~\bibnamefont {Garbarino}},\ and\ \bibinfo
  {author} {\bibfnamefont {J.}~\bibnamefont {Hlinka}},\ }\href
  {https://doi.org/10.1088/0953-8984/25/11/115901} {\bibfield  {journal}
  {\bibinfo  {journal} {J. Phys.: Condens. Matter}\ }\textbf {\bibinfo {volume}
  {25}},\ \bibinfo {pages} {115901} (\bibinfo {year} {2013})}\BibitemShut
  {NoStop}%
\bibitem [{\citenamefont {Zamaraite}\ \emph {et~al.}(2020)\citenamefont
  {Zamaraite}, \citenamefont {Liubachko}, \citenamefont {Yevych}, \citenamefont
  {Oleaga}, \citenamefont {Salazar}, \citenamefont {Dziaugys}, \citenamefont
  {Banys},\ and\ \citenamefont {Vysochanskii}}]{Zamaraite2020}%
  \BibitemOpen
  \bibfield  {author} {\bibinfo {author} {\bibfnamefont {I.}~\bibnamefont
  {Zamaraite}}, \bibinfo {author} {\bibfnamefont {V.}~\bibnamefont
  {Liubachko}}, \bibinfo {author} {\bibfnamefont {R.}~\bibnamefont {Yevych}},
  \bibinfo {author} {\bibfnamefont {A.}~\bibnamefont {Oleaga}}, \bibinfo
  {author} {\bibfnamefont {A.}~\bibnamefont {Salazar}}, \bibinfo {author}
  {\bibfnamefont {A.}~\bibnamefont {Dziaugys}}, \bibinfo {author}
  {\bibfnamefont {J.}~\bibnamefont {Banys}},\ and\ \bibinfo {author}
  {\bibfnamefont {Y.}~\bibnamefont {Vysochanskii}},\ }\href
  {https://doi.org/10.1063/5.0009762} {\bibfield  {journal} {\bibinfo
  {journal} {J. Appl. Phys.}\ }\textbf {\bibinfo {volume} {128}},\ \bibinfo
  {pages} {234105} (\bibinfo {year} {2020})}\BibitemShut {NoStop}%
\bibitem [{\citenamefont {Shchennikov}\ \emph {et~al.}(2011)\citenamefont
  {Shchennikov}, \citenamefont {Morozova}, \citenamefont {Tyagur},
  \citenamefont {Tyagur},\ and\ \citenamefont {Ovsyannikov}}]{shchennikov2011}%
  \BibitemOpen
  \bibfield  {author} {\bibinfo {author} {\bibfnamefont {V.~V.}\ \bibnamefont
  {Shchennikov}}, \bibinfo {author} {\bibfnamefont {N.~V.}\ \bibnamefont
  {Morozova}}, \bibinfo {author} {\bibfnamefont {I.}~\bibnamefont {Tyagur}},
  \bibinfo {author} {\bibfnamefont {Y.}~\bibnamefont {Tyagur}},\ and\ \bibinfo
  {author} {\bibfnamefont {S.~V.}\ \bibnamefont {Ovsyannikov}},\ }\href
  {https://doi.org/10.1063/1.3662926} {\bibfield  {journal} {\bibinfo
  {journal} {Applied Physics Letters}\ }\textbf {\bibinfo {volume} {99}},\
  \bibinfo {pages} {212104} (\bibinfo {year} {2011})}\BibitemShut {NoStop}%
\bibitem [{\citenamefont {Ovsyannikov}\ \emph {et~al.}(2013)\citenamefont
  {Ovsyannikov}, \citenamefont {Gou}, \citenamefont {Morozova}, \citenamefont
  {Tyagur}, \citenamefont {Tyagur},\ and\ \citenamefont
  {Shchennikov}}]{ovsyannikov2013}%
  \BibitemOpen
  \bibfield  {author} {\bibinfo {author} {\bibfnamefont {S.~V.}\ \bibnamefont
  {Ovsyannikov}}, \bibinfo {author} {\bibfnamefont {H.}~\bibnamefont {Gou}},
  \bibinfo {author} {\bibfnamefont {N.~V.}\ \bibnamefont {Morozova}}, \bibinfo
  {author} {\bibfnamefont {I.}~\bibnamefont {Tyagur}}, \bibinfo {author}
  {\bibfnamefont {Y.}~\bibnamefont {Tyagur}},\ and\ \bibinfo {author}
  {\bibfnamefont {V.~V.}\ \bibnamefont {Shchennikov}},\ }\href
  {https://doi.org/10.1063/1.4772624} {\bibfield  {journal} {\bibinfo
  {journal} {Journal of Applied Physics}\ }\textbf {\bibinfo {volume} {113}},\
  \bibinfo {pages} {013511} (\bibinfo {year} {2013})}\BibitemShut {NoStop}%
\bibitem [{\citenamefont {Ovsyannikov}\ \emph {et~al.}(2017)\citenamefont
  {Ovsyannikov}, \citenamefont {Morozova}, \citenamefont {Korobeinikov},
  \citenamefont {Haborets}, \citenamefont {Yevych}, \citenamefont
  {Vysochanskii},\ and\ \citenamefont {Shchennikov}}]{ovsyannikov2017}%
  \BibitemOpen
  \bibfield  {author} {\bibinfo {author} {\bibfnamefont {S.~V.}\ \bibnamefont
  {Ovsyannikov}}, \bibinfo {author} {\bibfnamefont {N.~V.}\ \bibnamefont
  {Morozova}}, \bibinfo {author} {\bibfnamefont {I.~V.}\ \bibnamefont
  {Korobeinikov}}, \bibinfo {author} {\bibfnamefont {V.}~\bibnamefont
  {Haborets}}, \bibinfo {author} {\bibfnamefont {R.}~\bibnamefont {Yevych}},
  \bibinfo {author} {\bibfnamefont {Y.}~\bibnamefont {Vysochanskii}},\ and\
  \bibinfo {author} {\bibfnamefont {V.~V.}\ \bibnamefont {Shchennikov}},\
  }\href {https://doi.org/10.1039/C6DT03854A} {\bibfield  {journal} {\bibinfo
  {journal} {Dalton Transactions}\ }\textbf {\bibinfo {volume} {46}},\ \bibinfo
  {pages} {4245} (\bibinfo {year} {2017})}\BibitemShut {NoStop}%
\bibitem [{\citenamefont {Yue}\ \emph {et~al.}(2023)\citenamefont {Yue},
  \citenamefont {Zhong}, \citenamefont {Wen}, \citenamefont {Wang},
  \citenamefont {Yu}, \citenamefont {Yu}, \citenamefont {Chen}, \citenamefont
  {Wang},\ and\ \citenamefont {Hong}}]{yue2021}%
  \BibitemOpen
  \bibfield  {author} {\bibinfo {author} {\bibfnamefont {B.}~\bibnamefont
  {Yue}}, \bibinfo {author} {\bibfnamefont {W.}~\bibnamefont {Zhong}}, \bibinfo
  {author} {\bibfnamefont {T.}~\bibnamefont {Wen}}, \bibinfo {author}
  {\bibfnamefont {Y.}~\bibnamefont {Wang}}, \bibinfo {author} {\bibfnamefont
  {H.}~\bibnamefont {Yu}}, \bibinfo {author} {\bibfnamefont {X.}~\bibnamefont
  {Yu}}, \bibinfo {author} {\bibfnamefont {C.}~\bibnamefont {Chen}}, \bibinfo
  {author} {\bibfnamefont {J.-T.}\ \bibnamefont {Wang}},\ and\ \bibinfo
  {author} {\bibfnamefont {F.}~\bibnamefont {Hong}},\ }\href
  {https://doi.org/10.1103/PhysRevB.107.L140501} {\bibfield  {journal}
  {\bibinfo  {journal} {Phys. Rev. B}\ }\textbf {\bibinfo {volume} {107}},\
  \bibinfo {pages} {L140501} (\bibinfo {year} {2023})}\BibitemShut {NoStop}%
\bibitem [{\citenamefont {Jia}\ \emph {et~al.}(2011)\citenamefont {Jia},
  \citenamefont {Jiramongkolchai}, \citenamefont {Suchomel}, \citenamefont
  {Toby}, \citenamefont {Checkelsky}, \citenamefont {Ong},\ and\ \citenamefont
  {Cava}}]{jia2011}%
  \BibitemOpen
  \bibfield  {author} {\bibinfo {author} {\bibfnamefont {S.}~\bibnamefont
  {Jia}}, \bibinfo {author} {\bibfnamefont {P.}~\bibnamefont
  {Jiramongkolchai}}, \bibinfo {author} {\bibfnamefont {M.~R.}\ \bibnamefont
  {Suchomel}}, \bibinfo {author} {\bibfnamefont {B.~H.}\ \bibnamefont {Toby}},
  \bibinfo {author} {\bibfnamefont {J.~G.}\ \bibnamefont {Checkelsky}},
  \bibinfo {author} {\bibfnamefont {N.~P.}\ \bibnamefont {Ong}},\ and\ \bibinfo
  {author} {\bibfnamefont {R.}~\bibnamefont {Cava}},\ }\href
  {https://doi.org/10.1038/nphys1868} {\bibfield  {journal} {\bibinfo
  {journal} {Nature Physics}\ }\textbf {\bibinfo {volume} {7}},\ \bibinfo
  {pages} {207} (\bibinfo {year} {2011})}\BibitemShut {NoStop}%
\bibitem [{\citenamefont {Imai}\ \emph {et~al.}(2014)\citenamefont {Imai},
  \citenamefont {Michioka}, \citenamefont {Ohta}, \citenamefont {Matsuo},
  \citenamefont {Kindo}, \citenamefont {Ueda},\ and\ \citenamefont
  {Yoshimura}}]{imai2014}%
  \BibitemOpen
  \bibfield  {author} {\bibinfo {author} {\bibfnamefont {M.}~\bibnamefont
  {Imai}}, \bibinfo {author} {\bibfnamefont {C.}~\bibnamefont {Michioka}},
  \bibinfo {author} {\bibfnamefont {H.}~\bibnamefont {Ohta}}, \bibinfo {author}
  {\bibfnamefont {A.}~\bibnamefont {Matsuo}}, \bibinfo {author} {\bibfnamefont
  {K.}~\bibnamefont {Kindo}}, \bibinfo {author} {\bibfnamefont
  {H.}~\bibnamefont {Ueda}},\ and\ \bibinfo {author} {\bibfnamefont
  {K.}~\bibnamefont {Yoshimura}},\ }\href
  {https://doi.org/10.1103/PhysRevB.90.014407} {\bibfield  {journal} {\bibinfo
  {journal} {Phys. Rev. B}\ }\textbf {\bibinfo {volume} {90}},\ \bibinfo
  {pages} {014407} (\bibinfo {year} {2014})}\BibitemShut {NoStop}%
\bibitem [{\citenamefont {Chen}\ \emph {et~al.}(2022)\citenamefont {Chen},
  \citenamefont {Wang}, \citenamefont {Ying}, \citenamefont {Huang},
  \citenamefont {Gou}, \citenamefont {Zhang}, \citenamefont {Li}, \citenamefont
  {Hosono}, \citenamefont {Guo},\ and\ \citenamefont {Chen}}]{chen2022}%
  \BibitemOpen
  \bibfield  {author} {\bibinfo {author} {\bibfnamefont {X.}~\bibnamefont
  {Chen}}, \bibinfo {author} {\bibfnamefont {J.}~\bibnamefont {Wang}}, \bibinfo
  {author} {\bibfnamefont {T.}~\bibnamefont {Ying}}, \bibinfo {author}
  {\bibfnamefont {D.}~\bibnamefont {Huang}}, \bibinfo {author} {\bibfnamefont
  {H.}~\bibnamefont {Gou}}, \bibinfo {author} {\bibfnamefont {Q.}~\bibnamefont
  {Zhang}}, \bibinfo {author} {\bibfnamefont {Y.}~\bibnamefont {Li}}, \bibinfo
  {author} {\bibfnamefont {H.}~\bibnamefont {Hosono}}, \bibinfo {author}
  {\bibfnamefont {J.-g.}\ \bibnamefont {Guo}},\ and\ \bibinfo {author}
  {\bibfnamefont {X.}~\bibnamefont {Chen}},\ }\href
  {https://doi.org/10.1103/PhysRevB.106.184502} {\bibfield  {journal} {\bibinfo
   {journal} {Phys. Rev. B}\ }\textbf {\bibinfo {volume} {106}},\ \bibinfo
  {pages} {184502} (\bibinfo {year} {2022})}\BibitemShut {NoStop}%
\bibitem [{\citenamefont {Oganov}\ and\ \citenamefont {Glass}(2006)}]{uspex_1}%
  \BibitemOpen
  \bibfield  {author} {\bibinfo {author} {\bibfnamefont {A.~R.}\ \bibnamefont
  {Oganov}}\ and\ \bibinfo {author} {\bibfnamefont {C.~W.}\ \bibnamefont
  {Glass}},\ }\href {https://doi.org/10.1063/1.2210932} {\bibfield  {journal}
  {\bibinfo  {journal} {The Journal of Chemical Physics}\ }\textbf {\bibinfo
  {volume} {124}},\ \bibinfo {eid} {244704} (\bibinfo {year}
  {2006})}\BibitemShut {NoStop}%
\bibitem [{\citenamefont {Oganov}\ \emph {et~al.}(2011)\citenamefont {Oganov},
  \citenamefont {Lyakhov},\ and\ \citenamefont {Valle}}]{uspex_2}%
  \BibitemOpen
  \bibfield  {author} {\bibinfo {author} {\bibfnamefont {A.~R.}\ \bibnamefont
  {Oganov}}, \bibinfo {author} {\bibfnamefont {A.~O.}\ \bibnamefont
  {Lyakhov}},\ and\ \bibinfo {author} {\bibfnamefont {M.}~\bibnamefont
  {Valle}},\ }\href {https://doi.org/10.1021/ar1001318} {\bibfield  {journal}
  {\bibinfo  {journal} {Accounts of Chemical Research}\ }\textbf {\bibinfo
  {volume} {44}},\ \bibinfo {pages} {227} (\bibinfo {year} {2011})}\BibitemShut
  {NoStop}%
\bibitem [{\citenamefont {Lyakhov}\ \emph {et~al.}(2013)\citenamefont
  {Lyakhov}, \citenamefont {Oganov}, \citenamefont {Stokes},\ and\
  \citenamefont {Zhu}}]{uspex_3}%
  \BibitemOpen
  \bibfield  {author} {\bibinfo {author} {\bibfnamefont {A.~O.}\ \bibnamefont
  {Lyakhov}}, \bibinfo {author} {\bibfnamefont {A.~R.}\ \bibnamefont {Oganov}},
  \bibinfo {author} {\bibfnamefont {H.~T.}\ \bibnamefont {Stokes}},\ and\
  \bibinfo {author} {\bibfnamefont {Q.}~\bibnamefont {Zhu}},\ }\href
  {https://doi.org/10.1016/j.cpc.2012.12.009} {\bibfield  {journal} {\bibinfo
  {journal} {Computer Physics Communications}\ }\textbf {\bibinfo {volume}
  {184}},\ \bibinfo {pages} {1172 } (\bibinfo {year} {2013})}\BibitemShut
  {NoStop}%
\bibitem [{\citenamefont {Kresse}\ and\ \citenamefont
  {Hafner}(1993)}]{VASP_Kresse:1993}%
  \BibitemOpen
  \bibfield  {author} {\bibinfo {author} {\bibfnamefont {G.}~\bibnamefont
  {Kresse}}\ and\ \bibinfo {author} {\bibfnamefont {J.}~\bibnamefont
  {Hafner}},\ }\href {https://doi.org/10.1103/PhysRevB.47.558} {\bibfield
  {journal} {\bibinfo  {journal} {Phys. Rev. B}\ }\textbf {\bibinfo {volume}
  {47}},\ \bibinfo {pages} {558} (\bibinfo {year} {1993})}\BibitemShut
  {NoStop}%
\bibitem [{\citenamefont {Kresse}\ and\ \citenamefont
  {Furthm\"uller}(1996)}]{VASP_Kresse:1996}%
  \BibitemOpen
  \bibfield  {author} {\bibinfo {author} {\bibfnamefont {G.}~\bibnamefont
  {Kresse}}\ and\ \bibinfo {author} {\bibfnamefont {J.}~\bibnamefont
  {Furthm\"uller}},\ }\href {https://doi.org/10.1103/PhysRevB.54.11169}
  {\bibfield  {journal} {\bibinfo  {journal} {Phys. Rev. B}\ }\textbf {\bibinfo
  {volume} {54}},\ \bibinfo {pages} {11169} (\bibinfo {year}
  {1996})}\BibitemShut {NoStop}%
\bibitem [{\citenamefont {Perdew}\ and\ \citenamefont {Wang}(1992)}]{LDA}%
  \BibitemOpen
  \bibfield  {author} {\bibinfo {author} {\bibfnamefont {J.~P.}\ \bibnamefont
  {Perdew}}\ and\ \bibinfo {author} {\bibfnamefont {Y.}~\bibnamefont {Wang}},\
  }\href {https://doi.org/10.1103/PhysRevB.45.13244} {\bibfield  {journal}
  {\bibinfo  {journal} {Phys. Rev. B}\ }\textbf {\bibinfo {volume} {45}},\
  \bibinfo {pages} {13244} (\bibinfo {year} {1992})}\BibitemShut {NoStop}%
\bibitem [{\citenamefont {Sun}\ \emph {et~al.}(2015)\citenamefont {Sun},
  \citenamefont {Ruzsinszky},\ and\ \citenamefont {Perdew}}]{SCAN}%
  \BibitemOpen
  \bibfield  {author} {\bibinfo {author} {\bibfnamefont {J.}~\bibnamefont
  {Sun}}, \bibinfo {author} {\bibfnamefont {A.}~\bibnamefont {Ruzsinszky}},\
  and\ \bibinfo {author} {\bibfnamefont {J.~P.}\ \bibnamefont {Perdew}},\
  }\href {https://doi.org/10.1103/PhysRevLett.115.036402} {\bibfield  {journal}
  {\bibinfo  {journal} {Phys. Rev. Lett.}\ }\textbf {\bibinfo {volume} {115}},\
  \bibinfo {pages} {036402} (\bibinfo {year} {2015})}\BibitemShut {NoStop}%
\bibitem [{\citenamefont {Haborets}\ \emph {et~al.}(2019)\citenamefont
  {Haborets}, \citenamefont {Rushchanskii}, \citenamefont {Medulych},
  \citenamefont {Glazyrin},\ and\ \citenamefont
  {Vysochanskii}}]{conference2019}%
  \BibitemOpen
  \bibfield  {author} {\bibinfo {author} {\bibfnamefont {V.}~\bibnamefont
  {Haborets}}, \bibinfo {author} {\bibfnamefont {K.}~\bibnamefont
  {Rushchanskii}}, \bibinfo {author} {\bibfnamefont {M.}~\bibnamefont
  {Medulych}}, \bibinfo {author} {\bibfnamefont {K.}~\bibnamefont {Glazyrin}},\
  and\ \bibinfo {author} {\bibfnamefont {Y.}~\bibnamefont {Vysochanskii}},\
  }in\ \href@noop {} {\emph {\bibinfo {booktitle} {Materials of the
  School-conference of young scientists «Modern Material Science: Physics,
  Chemistry, Technology (MMSPCT-2019)», Uzhgorod Vodogray Ukraine, 27 - 31 May
  2019}}},\ \bibinfo {editor} {edited by\ \bibinfo {editor} {\bibfnamefont
  {A.~H.}\ \bibnamefont {Naumovets'}}}\ (\bibinfo  {publisher} {PE Sabov
  A.M.},\ \bibinfo {address} {Uzhgorod},\ \bibinfo {year} {2019})\ pp.\
  \bibinfo {pages} {50--53}\BibitemShut {NoStop}%
\bibitem [{\citenamefont {Chakraverty}(1979)}]{chakraverty1979}%
  \BibitemOpen
  \bibfield  {author} {\bibinfo {author} {\bibfnamefont {B.}~\bibnamefont
  {Chakraverty}},\ }\href {https://doi.org/10.1051/jphyslet:0197900400509900}
  {\bibfield  {journal} {\bibinfo  {journal} {{Journal de Physique Lettres}}\
  }\textbf {\bibinfo {volume} {40}},\ \bibinfo {pages} {99} (\bibinfo {year}
  {1979})}\BibitemShut {NoStop}%
\bibitem [{\citenamefont {M\"uller}\ and\ \citenamefont
  {Burkard}(1979)}]{PhysRevB.19.3593}%
  \BibitemOpen
  \bibfield  {author} {\bibinfo {author} {\bibfnamefont {K.~A.}\ \bibnamefont
  {M\"uller}}\ and\ \bibinfo {author} {\bibfnamefont {H.}~\bibnamefont
  {Burkard}},\ }\href {https://doi.org/10.1103/PhysRevB.19.3593} {\bibfield
  {journal} {\bibinfo  {journal} {Phys. Rev. B}\ }\textbf {\bibinfo {volume}
  {19}},\ \bibinfo {pages} {3593} (\bibinfo {year} {1979})}\BibitemShut
  {NoStop}%
\bibitem [{\citenamefont {Micnas}\ \emph {et~al.}(1988)\citenamefont {Micnas},
  \citenamefont {Ranninger},\ and\ \citenamefont
  {Robaszkiewicz}}]{micnas:jpa-00229287}%
  \BibitemOpen
  \bibfield  {author} {\bibinfo {author} {\bibfnamefont {R.}~\bibnamefont
  {Micnas}}, \bibinfo {author} {\bibfnamefont {J.}~\bibnamefont {Ranninger}},\
  and\ \bibinfo {author} {\bibfnamefont {S.}~\bibnamefont {Robaszkiewicz}},\
  }\href {https://doi.org/10.1051/jphyscol:19888996} {\bibfield  {journal}
  {\bibinfo  {journal} {{Journal de Physique Colloques}}\ }\textbf {\bibinfo
  {volume} {49}},\ \bibinfo {pages} {C8} (\bibinfo {year} {1988})}\BibitemShut
  {NoStop}%
\bibitem [{\citenamefont {Varma}(1988)}]{PhysRevLett.61.2713}%
  \BibitemOpen
  \bibfield  {author} {\bibinfo {author} {\bibfnamefont {C.~M.}\ \bibnamefont
  {Varma}},\ }\href {https://doi.org/10.1103/PhysRevLett.61.2713} {\bibfield
  {journal} {\bibinfo  {journal} {Phys. Rev. Lett.}\ }\textbf {\bibinfo
  {volume} {61}},\ \bibinfo {pages} {2713} (\bibinfo {year}
  {1988})}\BibitemShut {NoStop}%
\bibitem [{\citenamefont {Mott}(1993)}]{Mott1993}%
  \BibitemOpen
  \bibfield  {author} {\bibinfo {author} {\bibfnamefont {N.}~\bibnamefont
  {Mott}},\ }\href {https://doi.org/10.1016/0378-4371(93)9051} {\bibfield
  {journal} {\bibinfo  {journal} {Physica A: Statistical Mechanics and its
  Applications}\ }\textbf {\bibinfo {volume} {200}},\ \bibinfo {pages} {127}
  (\bibinfo {year} {1993})}\BibitemShut {NoStop}%
\bibitem [{\citenamefont {Rowley}\ \emph {et~al.}(2014)\citenamefont {Rowley},
  \citenamefont {Spalek}, \citenamefont {Smith}, \citenamefont {Dean},
  \citenamefont {Itoh}, \citenamefont {Scott}, \citenamefont {Lonzarich},\ and\
  \citenamefont {Saxena}}]{Rowley_2014}%
  \BibitemOpen
  \bibfield  {author} {\bibinfo {author} {\bibfnamefont {S.~E.}\ \bibnamefont
  {Rowley}}, \bibinfo {author} {\bibfnamefont {L.~J.}\ \bibnamefont {Spalek}},
  \bibinfo {author} {\bibfnamefont {R.~P.}\ \bibnamefont {Smith}}, \bibinfo
  {author} {\bibfnamefont {M.~P.~M.}\ \bibnamefont {Dean}}, \bibinfo {author}
  {\bibfnamefont {M.}~\bibnamefont {Itoh}}, \bibinfo {author} {\bibfnamefont
  {J.~F.}\ \bibnamefont {Scott}}, \bibinfo {author} {\bibfnamefont {G.~G.}\
  \bibnamefont {Lonzarich}},\ and\ \bibinfo {author} {\bibfnamefont {S.~S.}\
  \bibnamefont {Saxena}},\ }\href {https://doi.org/10.1038/nphys2924}
  {\bibfield  {journal} {\bibinfo  {journal} {Nature Physics}\ }\textbf
  {\bibinfo {volume} {10}},\ \bibinfo {pages} {367–372} (\bibinfo {year}
  {2014})}\BibitemShut {NoStop}%
\bibitem [{\citenamefont {Volkov}\ \emph {et~al.}(2022)\citenamefont {Volkov},
  \citenamefont {Chandra},\ and\ \citenamefont {Coleman}}]{Volkov_2022}%
  \BibitemOpen
  \bibfield  {author} {\bibinfo {author} {\bibfnamefont {P.~A.}\ \bibnamefont
  {Volkov}}, \bibinfo {author} {\bibfnamefont {P.}~\bibnamefont {Chandra}},\
  and\ \bibinfo {author} {\bibfnamefont {P.}~\bibnamefont {Coleman}},\
  }\bibfield  {journal} {\bibinfo  {journal} {Nature Communications}\ }\textbf
  {\bibinfo {volume} {13}},\ \href {https://doi.org/10.1038/s41467-022-32303-2}
  {10.1038/s41467-022-32303-2} (\bibinfo {year} {2022})\BibitemShut {NoStop}%
\bibitem [{\citenamefont {{J\"{u}lich Supercomputing Centre}}(2018)}]{jureca}%
  \BibitemOpen
  \bibfield  {author} {\bibinfo {author} {\bibnamefont {{J\"{u}lich
  Supercomputing Centre}}},\ }\href {https://doi.org/10.17815/jlsrf-4-121-1}
  {\bibfield  {journal} {\bibinfo  {journal} {Journal of large-scale research
  facilities}\ }\textbf {\bibinfo {volume} {4}},\ \bibinfo {pages} {A132}
  (\bibinfo {year} {2018})}\BibitemShut {NoStop}%
\bibitem [{\citenamefont {Momma}\ and\ \citenamefont {Izumi}(2011)}]{VESTA}%
  \BibitemOpen
  \bibfield  {author} {\bibinfo {author} {\bibfnamefont {K.}~\bibnamefont
  {Momma}}\ and\ \bibinfo {author} {\bibfnamefont {F.}~\bibnamefont {Izumi}},\
  }\href {https://doi.org/10.1107/S0021889811038970} {\bibfield  {journal}
  {\bibinfo  {journal} {Journal of Applied Crystallography}\ }\textbf {\bibinfo
  {volume} {44}},\ \bibinfo {pages} {1272} (\bibinfo {year}
  {2011})}\BibitemShut {NoStop}%
\bibitem [{\citenamefont {Stokes}\ and\ \citenamefont {Hatch}(2005)}]{FINDSYM}%
  \BibitemOpen
  \bibfield  {author} {\bibinfo {author} {\bibfnamefont {H.~T.}\ \bibnamefont
  {Stokes}}\ and\ \bibinfo {author} {\bibfnamefont {D.~M.}\ \bibnamefont
  {Hatch}},\ }\href {https://doi.org/10.1107/S0021889804031528} {\bibfield
  {journal} {\bibinfo  {journal} {Journal of Applied Crystallography}\ }\textbf
  {\bibinfo {volume} {38}},\ \bibinfo {pages} {237} (\bibinfo {year}
  {2005})}\BibitemShut {NoStop}%
\end{thebibliography}%

\end{document}